\documentclass[review]{elsarticle}
\usepackage{lineno,hyperref}
\usepackage{amsmath}

\usepackage[table,xcdraw]{xcolor}
\usepackage{diagbox}
\usepackage{subcaption}
\usepackage{tabularx}
\usepackage[T1]{fontenc}
\usepackage{makecell}
\usepackage{tipa}
\usepackage{multirow}

\journal{Journal of Computer Speech and Language}

\begin{document}

\begin{frontmatter}

\title{The Phonetic Footprint of Parkinson's Disease}

\author[fau]{Philipp Klumpp\corref{mycorrespondingauthor}}
\cortext[mycorrespondingauthor]{Corresponding author}
\ead{philipp.klumpp@fau.de}
\ead[url]{https://lme.tf.fau.de/person/klumpp/}
\author[fau,udea]{Tom{\'a}s Arias-Vergara}
\author[fau,udea]{Juan Camilo V{\'a}squez-Correa}
\author[fau,udea]{Paula Andrea P{\'e}rez-Toro}
\author[udea]{Juan Rafael Orozco-Arroyave}
\author[fau,ua]{Anton Batliner}
\author[fau]{Elmar N{\"o}th}

\address[fau]{Friedrich-Alexander-University Erlangen-Nuremberg, Pattern Recognition Lab, Martensstrasse 3, 91058 Erlangen, Germany}
\address[udea]{Universidad de Antioquia, Medell{\'i}n, Colombia}
\address[ua]{Chair of Embedded Intelligence for Health Care and Wellbeing, University of Augsburg, Germany}

\begin{abstract}
As one of the most prevalent neurodegenerative disorders, Parkinson's disease (PD) has a significant impact on the fine motor skills of patients. The complex interplay of different articulators during speech production and realization of required muscle tension become increasingly difficult, thus leading to a dysarthric speech. Characteristic patterns such as vowel instability, slurred pronunciation and slow speech can often be observed in the affected individuals and were analyzed in previous studies to determine the presence and progression of PD. In this work, we used a phonetic recognizer trained exclusively on healthy speech data to investigate how PD affected the phonetic footprint of patients. We rediscovered numerous patterns that had been described in previous contributions although our system had never seen any pathological speech previously. Furthermore, we could show that intermediate activations from the neural network could serve as feature vectors encoding information related to the disease state of individuals. We were also able to directly correlate the expert-rated intelligibility of a speaker with the mean confidence of phonetic predictions. Our results support the assumption that pathological data is not necessarily required to train systems that are capable of analyzing PD speech.
\end{abstract}


\end{frontmatter}

\section{Introduction}
Parkinson's disease (PD) is one of the most common neurological conditions~\cite{lee2016epidemiology} in aging societies. This neurodegenerative disorder is strongly characterized by its progressive motor symptoms which can be classified into four types: bradykinesia (slowness of movement), tremor (involuntary muscle activity), rigidity (freezing of gait) and postural instability~\cite{jankovic2008parkinson}. Whilst all these indications are increasingly visible in an individual as the degeneration of brain cells progresses, the ongoing deterioration has long become audible. The production of healthy speech requires a well-coordinated fine-motor interaction between the different articulators. Throughout the different stages of PD, fine-motor deficits lead to a number of speech impairments that are often referred to as Parkinsonian Dysarthria. It comprises various articulatory (slurred consonants), prosodic (monotonous pitch), phonatory (breathy voice) and respiratory (decreased loudness) symptoms~\cite{pinto2004treatments}. A computational assessment of speech signals can be performed to either classify the presence of PD or to monitor the progression of the ongoing neurological disorder. Previous studies used phonation and articulation features~\cite{almeida2019detecting}, voice onset time~\cite{arias2020automatic}, autoencoder representations~\cite{martinez2018deep} and other representations to detect the condition. Voice onset and offset features have also proven informative to track the progression of PD~\cite{orozco2016towards}. In a very recent study, the speech signal was utilized as a surrogate to estimate the medication state of patients~\cite{norel2020speech}. Acoustic models trained on large datasets from healthy speakers had already been used in previous studies~\cite{van2009speech} to predict the intelligibility of dysarthric speakers. Unlike the approach presented in this article, the described system required a phonetic reference to perform automatic speech alignment. The system described in \cite{christina2012hmm} was also trained exclusively with healthy speakers (seven male subjects). Despite the very small sample size, a correlation between the speech recognition performance and Frenchay Dysarthria Assessment scores could be reported. Training background models from larger datasets that only included healthy contributors was also performed for gait analysis~\cite{som2020unsupervised}. Robust features could be learned from the signals of healthy participants that were then used to classify PD. In a multi-modal study setup, the progression of PD was predicted with universal background models estimated from speech, handwriting and gait features of healthy subjects~\cite{vasquez2020comparison}.\\
A major weakness of most classification and regression (monitoring in the case of PD) methods is their lack of generalization when parameters have to be estimated from a small dataset. For modern deep architectures, the problem of small PD datasets can be dealt with by applying transfer learning methods using data from a secondary domain~\cite{naseer2020refining}. However, this solution is not optimal. The parameter-intensive feature extraction part could be trained with a large dataset from a different domain. Afterwards, these parameters are frozen and only the final layers would have to be estimated for the low-resource classification problem. Whilst this allows for a robust feature extraction, the following feature interpretation could be strongly affected by acoustic conditions, utilized hardware, signal pre-processing or study design, to mention a few.\\
Another important factor that naturally hinders generalization of machine learning models is the varying progression of the disease among patients. The presence and further development of symptoms is different for every individual~\cite{schrag2007rate}. This implies that any PD monitoring solution must be able to adapt to a patient-specific baseline instead of the one from a large study cohort.\\
With respect to study designs, most related works perform speech analysis on signals acquired from dedicated speech tasks, such as sustained vowels or diadochokinesis (DDK) tasks (rapid repetition of consonant-vowel clusters)~\cite{vasquez2019articulation,shahbakhi2014speech,skodda2013progression,rusz2013evaluation}. The major advantage of this setup is the combination of exercise and data collection. The main disadvantage is quite obvious as well: The trained model is bound to such speech exercises, hence unable to interpret free speech.\\
To overcome all the presented drawbacks, we propose the computation of phonetic footprints based on a fundamental acoustic analysis of speech signals. A phonetic footprint resembles the distribution of production probabilities among different phonemes or phonological classes for an individual speaker. The core idea behind the design of our recognizer was that it is trained only with speech samples from healthy speakers. This enabled us to include large general-purpose automatic speech recognition (ASR) datasets into the training procedure. To improve generalization even more and ensure that we are not learning strong language dependent patterns, we trained our model with independent datasets from three different languages. We used the final model to compute acoustic class probability densities for healthy subjects as well as PD patients to show how the phonetic footprint of speech signals changed between groups. Step by step we demonstrate how alterations in articulation are conserved from speech exercises over read text all the way to free speech. Furthermore, we tried not only to identify changes in speech productions, but also to link these changes to the involved articulators.\\
The last hypothesis of this work arose from the nature of our training data. Our acoustic model has barely seen any dysarthric speech sample before. For a human listener, we know that the more dysarthric a speech sample sounds, the more unintelligible it becomes. We demonstrate that this perceived decrease of intelligibility was very well conserved in the results of our recognition model.\\
Phonetic footprints were not designed to serve as a basis for classifying PD. In fact, in a study about the evolution of PD diagnosis, only 11~\% of patients reported speech problems as one of the initial symptoms~\cite{jankovic2000evolution}. On the other side, more than three out of four patients reported tremor at the beginning of the disease. To mitigate PD symptoms, patients are being administered a pre-stage of the neurotransmitter dopamine. Only this pre-stage is able to pass the blood-brain barrier and make up for the lack of dopamine in the basal ganglia which was caused by the loss of dopaminergic brain cells~\cite{poewe2017parkinson}. The therapy success with levodopa has to be controlled on a regular basis, because at some point, the drug itself would cause strong motor symptoms which render the therapy useless~\cite{poewe2017parkinson}. An automatic and unobtrusive speech analysis could provide such supervision of therapy success on a daily basis and without requiring an expert clinician. Symptoms are developed differently by individual patients and their response to treatment with levodopa varies strongly~\cite{hacisalihzade1989optimization}. A phonetic footprint would have to be estimated individually for each patient to serve as a unique baseline. After the initial calibration, the footprint could be updated regularly with speech samples collected from different voice-user interfaces, such as smartphones or smart speakers. A larger amount of speech samples could also help to alleviate the influence of recognition errors on the final footprint. With the knowledge about how PD would affect the average footprint of a person as it progresses, it would be possible to search for distinct patterns, deviations from their individual baseline, to track if a patient responded well to therapy, and if not, whether this deviation was only "a bad day" or manifested over weeks. The study of phonetic footprints could ultimately improve PD monitoring in terms of availability, closer coverage and better adaption to the individual patient.\\
After a brief introduction into the different datasets used for training and PD analysis, we want to outline the multilingual concept of \textsc{phone} unions employed in this study. We then explain our acoustic model and how it was used to extract knowledge from unseen data. In the results section, we embed our findings with respect to different degrees of freedom in speech production, ranging from constricted speech exercises to free speech. We will then interpret the outcome of our experiments both with respect to the detected characteristics of PD speech as well as their progression throughout disease stages in the discussion. Finally, we want to highlight the key strengths of our method and we will point out remaining weak spots.

\section{Materials and Methods}

\begin{table}[t]
\small
\centering
\caption{Summary of training data distribution before and after augmentation.}
\label{tab:training_data}
\begin{tabular}{l|c|r|r}

\rowcolor[HTML]{C0C0C0}
Language & Gender [f\,/\,m] & Hours before Aug. & Hours after Aug. \\ \hline
German	&	307\,/\,286	&	29.1    & 58.2	\\ \hline
English &  56\,/\,112	&	5.3 &   10.6	\\ \hline
Mexican Spanish	&	51\,/\,49	&	6.2 & 12.4	\\ \hline \hline
\textbf{Total}	&	\textbf{414\,/\,447}	&	\textbf{40.6}    &   \textbf{81.2}	\\
\end{tabular}
\end{table}

\subsection{Training data}
Our acoustic model was trained with German, English and Spanish ASR datasets. They were carefully selected due to their high quality in annotation and acoustic properties. The high annotation quality would result in fewer phonetic label errors, thus giving us a more reliable acoustic model. The PD dataset was recorded in a noise-free environment with high-quality microphones. We therefore tried to train our acoustic models with data of similar acoustic properties. We used a subset of the German Verbmobil corpus~\cite{wahlster2013verbmobil} containing around 29 hours of dialogue speech recordings from 593 speakers (307 female, 286 male). This was the largest of the three datasets. For our purpose we downsampled the data to 16 kHz using 16 bits/sample and mono-channel configuration. We distributed the speakers randomly into training (26.1~h), development (1.5~h) and test (1.5~h) sets. Phonetic segmentations were created by forced-alignment with the Kaldi~\cite{povey2011kaldi} ASR toolkit.\\
We also incorporated the TIMIT Acoustic-Phonetic Continuous Speech Corpus \cite{garofolo1993darpa}. It is composed of American English speech samples collected from a total of 630 speakers, 438 (70\%) male and 192 (30\%) female, from eight major dialect regions. Every speaker contributed ten recordings of phonetically rich text reading tasks. The PCM audio files had been recorded with a resolution of 16 bits at a sampling rate of 16~kHz and mono-channel configuration. The TIMIT corpus was hand-labeled to provide time-aligned segmentation at the orthographic, phonetic and word level. It is distributed with a predefined split into training (3.9~h) and test (1.4~h) sets. 168 of the original 630 speakers were part of the test subset, 112 (67\%) male and 56 (33\%) female. During training, we used a portion of 5\% of all speakers from the training split for validation.\\
Our PD dataset, which is described in subsection~\ref{subsec:PDdata}, was collected from native Colombian Spanish speakers. Therefore, we decided to include a Spanish corpus into the training as well. The DIMEx100 corpus~\cite{pineda2004dimex100} was recorded with 100 native Mexican Spanish speakers (49\% male, 51\% female). Every individual contributed 50 unique and 10 reference (identical for all speakers) phrases. Audio files were recorded in raw format at 44.1~kHz (once more we downsampled to 16~kHz), 16 bit resolution for a single channel. All speakers were distributed into training (5~h), development (0.6~h) and test set (0.6~h).\\
Every audio sample was used in its clean version and also with added Gaussian noise at different SNR levels (randomly chosen from 10~dB, 15~dB, 20~dB). The overall distribution of training data is summarized in Table~\ref{tab:training_data}.

\subsection{Parkinson's disease dataset}
\label{subsec:PDdata}
\begin{figure}[t]
  \centering
  \begin{subfigure}[b]{0.4\textwidth}
    \centering
    \includegraphics[width=\textwidth]{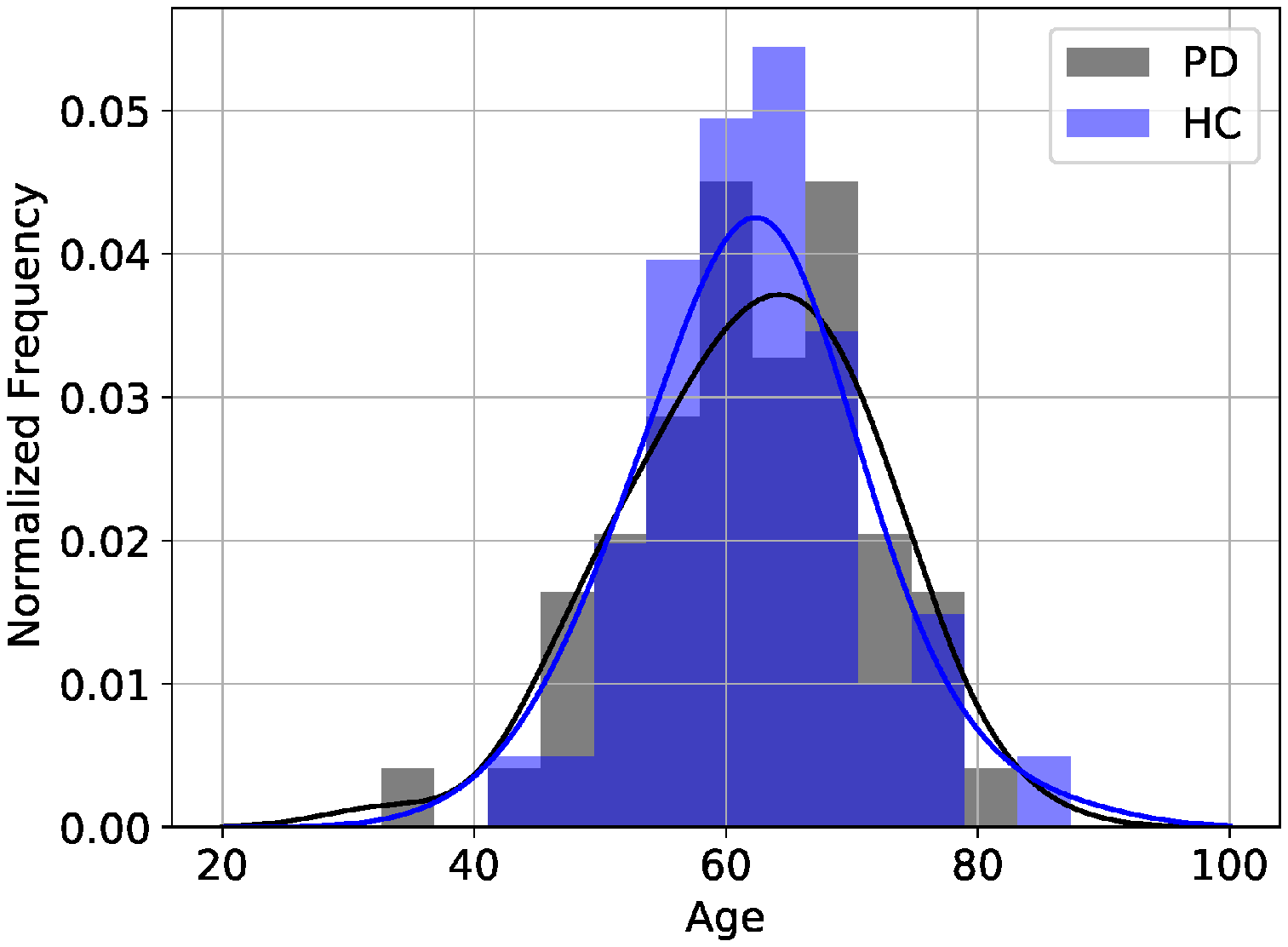}
    \caption{Age distribution.}
  \end{subfigure}
  \begin{subfigure}[b]{0.4\textwidth}
    \centering
    \includegraphics[width=\textwidth]{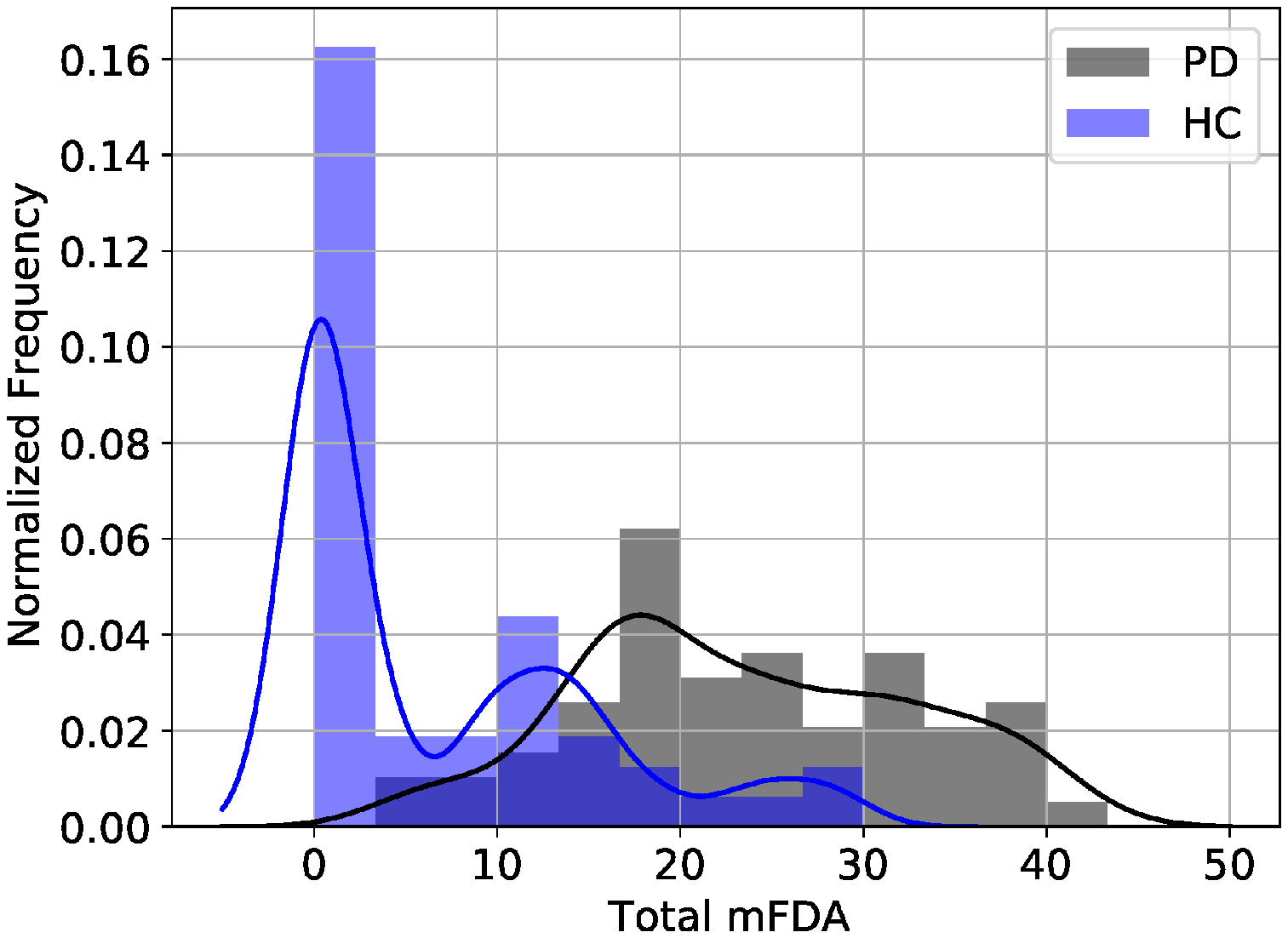}
    \caption{Distribution of mFDA scores.}
  \end{subfigure}
  \caption{Distribution of age and mFDA assessment scores for the groups of healthy speakers and Parkinson's patients.}
  \label{fig:data_description}
\end{figure}

In this study, we used the extended version of the PC-GITA corpus~\cite{vasquez2018towards} collected from native Colombian Spanish speakers. It consists of 68 pathological and 50 healthy control (HC) speakers. We reduced this distribution to those 58 (PD) and 48 (HC) who were recorded in a (portable) soundproof booth and had been assessed with respect to the modified Frenchay Dysarthria Assessment (mFDA)~\cite{vasquez2018towards}, an adapted version of the original FDA~\cite{enderby1980frenchay} for assessments from speech recordings. FDA is a common procedure to evaluate speech symptoms of PD patients. Unlike the very popular Unified Parkinson Disease Rating Scale (UPDRS)~\cite{goetz2008movement}, which mainly focuses on motor abilities, FDA only evaluates the oromotor abilities of a subject, thus providing a better impression of an individuals ability to articulate. Only including recordings from a noise-free environment was important to ensure equal acoustic conditions throughout our experiments. Every individual contributed samples of six diadochokinetic exercises, ten isolated sentences, one read text and one monologue. Audio signals were recorded in raw PCM, 16 bits resolution and mono-channel configuration at 16~kHz. Every participant was assessed by an expert therapist with respect to mFDA. The score aims to estimate an individual's articulatory abilities with respect to various items, namely the lips \mbox{[score range: 0\,-\,8]}, palate \mbox{[0\,-\,8]}, larynx \mbox{[0\,-\,12]}, respiration \mbox{[0\,-\,8]}, intelligibility \mbox{[0\,-\,4]}, monotonicity \mbox{[0\,-\,4]} and tongue \mbox{[0\,-\,8]}. This allows for a more precise localization of the morphological origin of a speech impairment inside the vocal tract. Throughout the whole mFDA value range \mbox{[0\,-\,52]}, larger values always indicate a stronger impairment.\\
The gender distribution among the HC group was 25 females and 23 males, among the PD patients it was 27 females, 31 males. The mean age was 62.0 (8.0 standard deviation) and 61.5 (9.3) for the HC and PD groups, respectively. The total mFDA score can be computed as a sum of all items described previously and was on average 6.6 (8.3) for the HC group and 22.7 (8.5) for the PD group. Figure~\ref{fig:data_description} provides an additional insight into the age and mFDA score distribution among healthy subjects and PD patients.

\subsection{Multilingual \textsc{phone} concept}
For any given language, there is a phoneme inventory (consonant system and vowel system) whose members represent those units that are both perceptually distinct and distinguish words in this language. One or more phones can be a realisation of the same phoneme; they are then called allophones or combinatorial variants.\\
For our multilingual acoustic model, we had to revise this concept as the term \textit{phoneme} is easily misused~\cite{Moore2019}. We instead decided to define a set of 35 \textsc{phone} unions (including one for silence), where every such \textsc{phone} would describe the set of elementary phones that could be used in German, English or Spanish to produce said \textsc{phone} class.\\
For some experiments, we also grouped \textsc{phones} into phonetic categories. This helped us interpret the overall capabilities of a subject with respect to certain classes of \textsc{phones}. We used a coarse and a fine set of phonetic classes. A full list of all phonetic categories is provided in Table~\ref{tab:phonetics}.

\begin{table}[t]
\centering
\caption{List of all 34 \textsc{phones} (excluding silence) used in this work, according to the international phonetic alphabet (IPA).}
\label{tab:phonetics}
\begin{tabular}{c|c|c}

\rowcolor[HTML]{C0C0C0}
\textsc{Phones} & Coarse phonetic class & Fine phonetic class  \\ \hline
l\,/\,j\,/\,w	&	Approximants	&	Approximants	\\ \hline
u\,/\,i\,/\,y &  Vowels	&	Closed Vowels	\\ \hline
e\,/\,o\,/\,\oe\,/\,\textepsilon\,/\,\textrevepsilon	&	Vowels	&	Mid-open Vowels	\\ \hline
a  	&	Vowels	&	Open Vowels	\\ \hline
s\,/\,\textesh\,/\,z\,/\,\textyogh 	&	Fricatives	&	Sibilant Fricatives	\\ \hline
\texttheta\,/\,\dh\,/\,h\,/\,v\,/\,x\,/\,f\,/\,\textgamma	&	Fricatives	&	Non-sibilant Fricatives	\\ \hline
n\,/\,m\,/\,\ng\,/\,\textltailn    	&	Nasals	&	Nasals	\\ \hline
r	&	Rhotics	&	Rhotics	\\ \hline
p\,/\,t\,/\,k	&	Stops	&	Unvoiced Stops	\\ \hline
b\,/\,d\,/\,g	&	Stops	&	Voiced Stops	\\
\end{tabular}
\end{table}

\subsection{Audio processing}
Every audio recording was first resampled to 16~kHz when necessary, followed by a root mean square normalization to a level of $-10$\,dB and removal of any DC offset. We then computed amplitude and phase spectrograms (2048 FFT points) over a window of 25~ms which was shifted by 5~ms. We found that including the phase information slightly improved the \textsc{phone} recognition results. Both spectrograms were converted to logarithmic scale of base 10 and afterwards filtered with a triangular Mel-bank~\cite{stevens1937scale} with 128 frequency bands to resemble the human perception of speech with a high resolution in the lower frequencies and a coarse resolution for high frequencies. The resulting dual-channel spectrograms were used throughout all experiments of this study.

\subsection{Deep recurrent \textsc{phone} recognition}
Our \textsc{phone} recognition model was comprised of two major components, a convolutional feature extraction part as well as a recurrent sequence analysis part. The whole neural network was trained in two separate steps to improve the final sequence prediction result. First, the network was optimized to perform framewise \textsc{phone} classification. At every time step $t$, the network would distribute a probability mass over 44 \textsc{phone} targets. Notice that for the first training stage, we had 9 additional \textsc{phone} targets. This was necessary because some of the \textsc{phone} labels of the described datasets were simply composites of other \textsc{phone} classes. For example, the ground truth class [\textteshlig] included in TIMIT is a composite of the \textsc{phones} [t] and [\textesh] that had already been defined. However, we could not just replace [\textteshlig] by the two underlying \textsc{phones} because we were missing the corresponding alignment information about where [t] ended and [\textesh] started. It would have been possible to compute new alignments for TIMIT, but then the exceptional quality of the dataset's manual phonetic annotation would have been lost.\\
A detailed description of the \textsc{phone} recognition model's architecture is provided in \ref{appendix:arch}. After we had successfully trained the framewise \textsc{phone} classifier, we slightly increased the parameter complexity of our recurrent network and retrained the recognizer, this time without alignment information. In other words, instead of telling the network about the explicit target \textsc{phone} at every time step, we used the connectionist temporal classification (CTC) loss function~\cite{graves2006connectionist} to predict an alignment-free \textsc{phone} sequence. This allowed us to disassemble the composite \textsc{phones} into their elementary components. Furthermore, framewise phonetic annotations are usually prone to label imprecision due to poor forced alignments or unclear \textsc{phone} boundaries. With the CTC loss, our model learned to identify inherent \textsc{phone} boundaries by itself.\\
All proposed architectures were trained with Adam optimizer~\cite{kingma2014adam} with initial learning rate of $0.001$. The learning rate decayed by a factor of $0.5$ if the validation loss did not improve for at least five subsequent epochs. One training batch contained 20 samples and each of the samples comprised a subsequence of 6 seconds duration chosen randomly over a recorded utterance. The core idea during pre-training was to learn a meaningful initialization of the convolutional feature extraction layers. After reaching convergence on this alignment task, we removed the two recurrent layers (200 hidden units) as well as the final classification layer for 44 \textsc{phone} targets and replaced them with two new recurrent layers (240 hidden units) and a classification layer projecting to the 35 sequential \textsc{phones}. During the framewise classification in the first step, we used cross entropy loss function. In the second stage, we used CTC loss function as described in~\cite{graves2006connectionist}. To improve overall model generalization, parameters were penalized by applying $L2$ regularization~\cite{DBLP:journals/corr/abs-1711-05101} with a rate of $0.0001$. During early epochs, we observed rather noisy validation loss values. One way of fixing this was to reduce the initial learning rate, but this would result in an overall slower training. Instead, we used gradient clipping to limit the $L2$ norm of gradients to a maximum value of $1.0$. This approach has proven to be beneficial for training other RNN architectures as well~\cite{graves2013generating}.\\
A model trained with CTC loss still performs a framewise prediction, estimating a probibility density function (PDF) over the total number of classes (35 in our case) and an additional helper class, commonly referred to as the blank label. A blank label indicates preservation of the previous state, meaning that the network's last emitted non-blank class is still present. This holds until a new non-blank class is predicted. For a given sequential input $X$ of length $T$, one can decode the sequence of predicted states $s$ of length $L \leq T$. The probability of observing path $p$ can be defined as

\begin{equation}
\label{eq:ctc_path_prob}
    P(p|X) = \prod_{t=1}^{T}y_c^t
\end{equation}

where $y_c^t$ denotes the probability of observing class $c$ at time $t$. The probability of any sequence $s$ can then be formulated as the sum of probabilities of all paths $p \in P_s$ that decode to $s$:

\begin{equation}
\label{eq:ctc_seq_prob}
    P(s|X) = \sum_{p \in P_s} P(p|X)
\end{equation}

Consequently, the most probable path $p$ might not always decode to the most probable sequence $s$. With a beam search decoding, it is possible to get a reliable estimate of $s$. To keep computational costs low, we decided to perform beam search with a beam width of 20, thus only keeping track of the 20 most probable \textsc{phone} sequences during decoding. It is common practice to extend equation~\ref{eq:ctc_seq_prob} with an additional term representing a phonetic language model. We decided not to include any such language dependent transition probabilities because they are likely to conflict with certain patterns of dedicated speech exercises that are uncommon in spoken language.\\
For our experiments, we used the trained \textsc{phone} recognition model to predict three types of information. The most important one was the \textsc{phone} sequence. It is alignment-free and contains a single \textsc{phone} symbol for every \textsc{phone} realization, independent of its duration. For every such sequence, we estimated each \textsc{phone}'s confidence by taking its posterior probability $y_s^t$ from the most probable path $p \in P_s$. Lastly, we also extracted the concatenated hidden states of the forward and backward passes from the last BiLSTM at the particular timestep $t$ as an intermediate, high-dimensional representation of this point in time.

\subsection{Methodology}
Although healthy speakers had also been assessed with respect to mFDA, they were never split into different sets according to their respective scores. The motivation was to compare pathological speech not only to unimpaired speech, but to speech from healthy subjects in general, among which pronunciation deficits could also occur for many other reasons.
\subsubsection{Speech exercises}
We first investigated speech productions from healthy and PD speakers collected during dedicated speech exercises. The exercise was split into three isolated tasks where an individual was asked to fluently repeat one of the syllables [pa], [ta] or [ka]. The group of PD patients was split into three sets according to their mFDA scores for lips (syllable [pa]) and tongue ([ta] and [ka]) such that the distribution among sets was as equal as possible. All sets used in the following were created with the goal to create the most equal distributions. We then investigated the phonetic capabilities of all groups by estimating the production probability of different \textsc{phones} or phonetic classes. Our hypothesis is that the decreasing muscle tension in PD patients and the resulting slurred pronunciation of stop sounds would leave a noticeable footprint in their phonetic profile for each task. Imprecisions in the production of \textsc{phones} [p], [t] and [k] would indicate how severely the involved articulators (lips and tongue, respectively) were affected. In all histograms showing relative frequencies of \textsc{phones} or phonetic groups, error bars indicate the standard deviation from the respective mean value.
\subsubsection{Intermediate network activations}
For the dedicated speech tasks, we also computed a 1-dimensional decomposition of intermediate RNN hidden states that were associated with one of the three stop sounds by performing a principal component analysis (PCA). A very similar analysis of intermediate neural network activations was successfully applied in a previous study as well~\cite{klumpp2020surgical}. For the sake of readability, we only split the PD group into two sets in this case. In the first step, all hidden states from the last RNN layer that were classified as one of the three unvoiced stop \textsc{phones} [p], [t] or [k], were collected from all 106 participant's [pa], [ta] and [ka] syllable repetitions. We then computed three one-dimensional PCAs, one for each collection of hidden states. After applying this PCA, we sorted the results according to mFDA scores to identify differences between hidden states of HC, mildly and severely affected patients. Lastly, we computed the mean confidence of \textsc{phone} predictions over all DDK exercises as a measure of perceived intelligibility of the \textsc{phone} recognition model, sorted according to the total mFDA score. This evaluation was also done for all following experiments.
\subsubsection{Isolated sentences}
In the next step, we performed a phonetic analysis of isolated sentences. Here, we did not evaluate with respect to each individual \textsc{phone} anymore, but rather evaluated the coarse and fine phonetic classes, because unlike for the speech tasks, we could not define any target \textsc{phones} anymore. The group of PD speakers was now divided into two sets, sorted according to their mFDA intelligibility rating. Notice that intelligibility resembles more the overall pronunciation capabilities of an individual, whereas the previously used items where dedicated to certain functional organs in the vocal tract that are involved in the production of a certain \textsc{phone}.
\subsubsection{Text reading and monologues}
The analyses of read texts (multiple sentences) and monologues (free speech) were done in the same way as for the isolated sentences. All explained differences were found to be statistically significant by performing t-tests and observing p-values $\ll$ 0.001.\\
To provide a better interpretation of the presented results, we performed a statistical analysis. We first determined all fine phonetic classes that were found to be significantly affected (as defined earlier in this subsection). For every class, we computed Spearman's correlation with the mFDA intelligibility score as well as the effect size (Cohen's \textit{d}) for every set as a measure of how strong each group deviated from the healthy reference. For the remaining three speech tasks, we chose the most affected coarse and fine phonetic class from the dedicated speech exercises and determined how correlation with intelligibility score and effect sizes changed. For the mean \textsc{phone} confidences, we computed Spearman's correlation with respect to total mFDA score and mFDA intelligibility. Additionally, we determined the effect sizes of the individual PD-sets and tasks with respect to the healthy reference. Whenever we computed Spearman's correlation, we always considered the assigned mFDA values of HC speakers as well to get an exact representation of how well a certain property (e.g. the mean prediction confidences) resembled this score. Statistical analysis was not performed on the PCA results because the intention behind this experiment was only to show that an intermediate feature vector can encode PD-related information.

\section{Results}

\subsection{\textsc{Phone} recognition}
Our framewise \textsc{phone} recognition model achieved a \textsc{phone} error rate (PER) of 18.3\%. The final CTC model reached a PER of 16.6\%. In this case, PER was computed as the ratio of deletions, insertions and replacements required to transform a predicted sequence to its ground truth. If we did not use the parameters from the framewise model's feature extraction layers as initialization (cold start) for the CTC model, we observed a slightly worse PER of 18.7\%.

\subsection{Phonetic analysis of Parkinson's Disease}

\begin{figure}[t]
  \centering
  \includegraphics[width=0.9\linewidth]{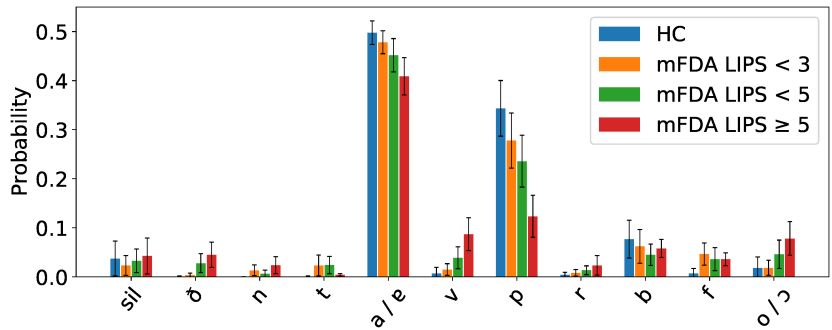}
  \caption{Mean posterior \textsc{phone} probabilities for the repetition of syllable [pa]. Error bars represent standard deviation from mean.}
  \label{fig:ddk4_full_p}
\end{figure}

\begin{figure}[t]
  \centering
  \includegraphics[width=0.9\linewidth]{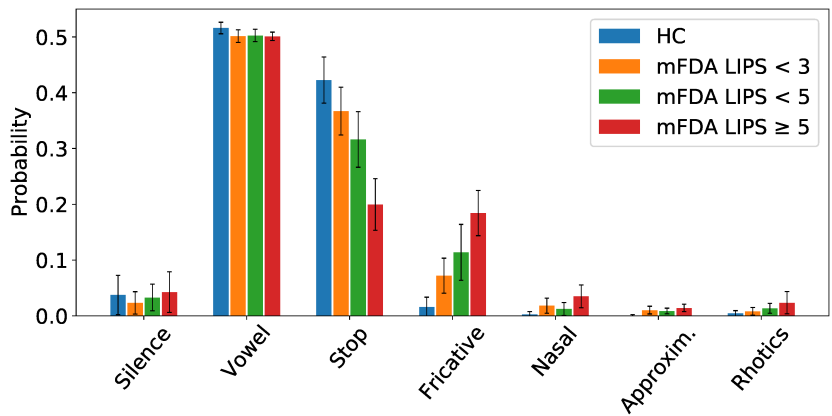}
  \caption{Mean posterior coarse phonetic class probabilities for the repetition of syllable [pa].}
  \label{fig:ddk4_cg_p}
\end{figure}

\subsubsection{Speech exercises}
Figures~\ref{fig:ddk4_full_p} and \ref{fig:ddk4_cg_p} show the mean posterior probabilities (MPP) of different \textsc{phones} (only \textsc{phones} where at least one set had an MPP > 2\%) and coarse phonetic classes for the repetition of syllable [pa], respectively. For the HC, the MPP for vowel [a] equals 49.8\%, which is very close to the expected value of 50\% if the syllable was repeated correctly. Notice that the expected value is not affected by the duration of \textsc{phone} productions, because the recognizer outputs a single \textsc{phone} symbol per realization. PD patients with an mFDA lips score below 3 achieved [a] MPP of 47.8\%. This value further decreased to 45.1\% in the second set for mFDA lips scores ranging from 3 to below 5. The last group with scores greater or equal to 5 only reached 40.9\%. Looking at the results in Figure~\ref{fig:ddk4_cg_p}, we observed that the MPP for vowels in general did not change much between the sets.\\
We found differences for the stop [p] to be even more pronounced. The MPP for HCs was 34.3\%. The patients with low, intermediate and high scores for the lips item achieved 27.8\%, 23.6\% and 12.3\% in MPP, respectively. Sometimes, instead of the unvoiced stop [p], we observed the voiced counterpart [b]. Such confusions could result from mispronunciations as well as from false recognition, because the two \textsc{phones} are very similar. For PD patients, we found that they have an increased probability of replacing the stop with fricatives such as [f], [v] or [\dh]. While healthy participants had an MPP of 1.6\% to produce a faulty fricative during the exercise, this value constantly increased along with the mFDA lips score, and reached a value of 18.4\% for the strongly impaired group.

\begin{figure}[t]
  \centering
  \includegraphics[width=0.9\linewidth]{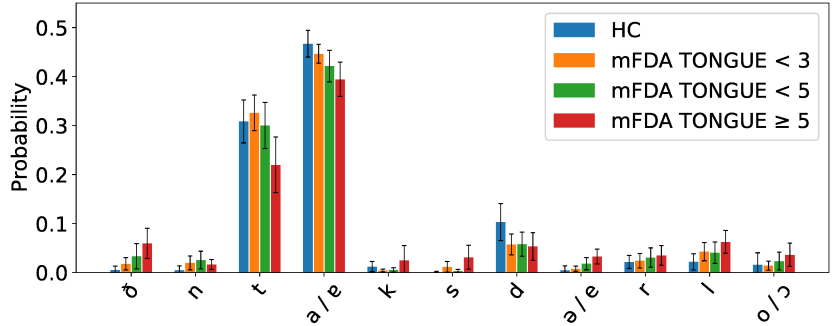}
  \caption{Mean posterior \textsc{phone} probabilities for the repetition of syllable [ta].}
  \label{fig:ddk5_full_t}
\end{figure}

\begin{figure}[t]
  \centering
  \includegraphics[width=0.9\linewidth]{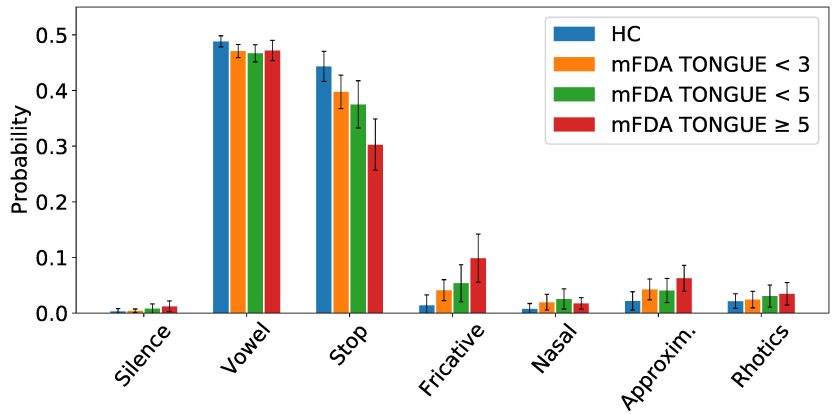}
  \caption{Mean posterior coarse phonetic group probabilities for the repetition of syllable [ta].}
  \label{fig:ddk5_cg_t}
\end{figure}

For the repetition of syllable [ta], the differences between sets appeared to be less pronounced compared to those of the first diadochokinesis (DDK) task. Figure~\ref{fig:ddk5_full_t} shows the MPPs for all relevant \textsc{phones} again. The set of mildly affected PD patients according to the mFDA score for the tongue item even achieved a slightly higher MPP for the stop [t]. In contrast to this, the HC group produced far more [d] stops, the voiced counterpart of [t]. Looking at the phonetic classes (Figure~\ref{fig:ddk5_cg_t}), the MPP of a stop sound in general (voiced or unvoiced) was still higher for the HC. When the stop [t] was not produced correctly, patients produced more fricatives and approximants at the same time. For the vowel [a], we observed the same results as before: The highest value for the HC group was 46.7\%, followed by a constant decline in MPP down to 39.4\% for patients with strong deficits associated with their tongue.

\begin{figure}[t]
  \centering
  \includegraphics[width=0.9\linewidth]{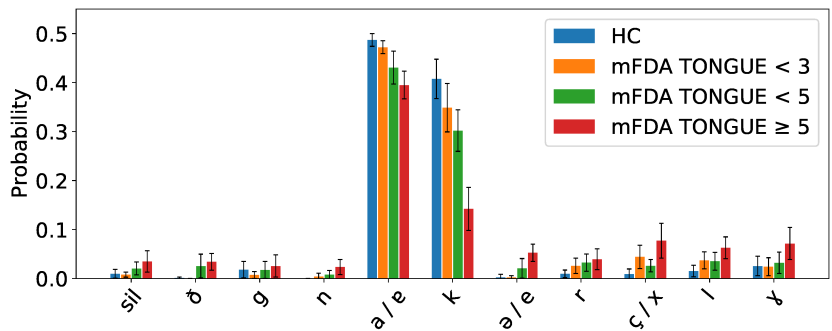}
  \caption{Mean posterior \textsc{phone} probabilities for the repetition of syllable [ka].}
  \label{fig:ddk6_full_k}
\end{figure}

\begin{figure}[t]
  \centering
  \includegraphics[width=0.9\linewidth]{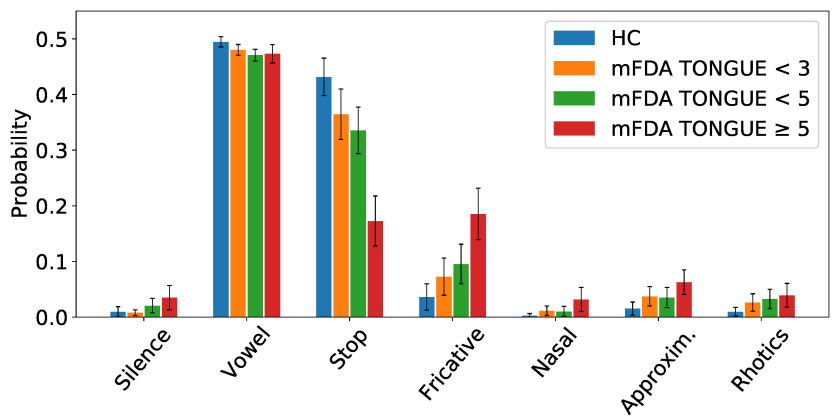}
  \caption{Mean posterior coarse phonetic group probabilities for the repetition of syllable [ka].}
  \label{fig:ddk6_cg_k}
\end{figure}

Results for the last DDK task (syllable [ka]) are presented in Figures~\ref{fig:ddk6_full_k} and \ref{fig:ddk6_cg_k}. The stability of vowel [a] showed the same characteristics as in the exercises before. The stop sound [k] had an MPP of 40.7\% for the HC. Mildly affected patients produced the correct stop with an MPP of 35.0\%, intermediates reached 30.2\%. The PD participants with an mFDA tongue score equal or greater than 5 only had an MPP of 14.2\%. Among the replacing \textsc{phones} were fricatives ([\dh], [x], [\textgamma]), rhotics ([r]) and approximants ([l]). For all three consonant-vowel-cluster exercises, we found that the standard deviations for the respective unvoiced and voiced stop \textsc{phones} were mostly higher than for the vowel. A similar pattern was observed for the coarse phonetic groups of stops and fricatives when compared to the low standard deviation for the vowels.

\begin{figure}[t]
  \centering
  \includegraphics[width=0.9\linewidth]{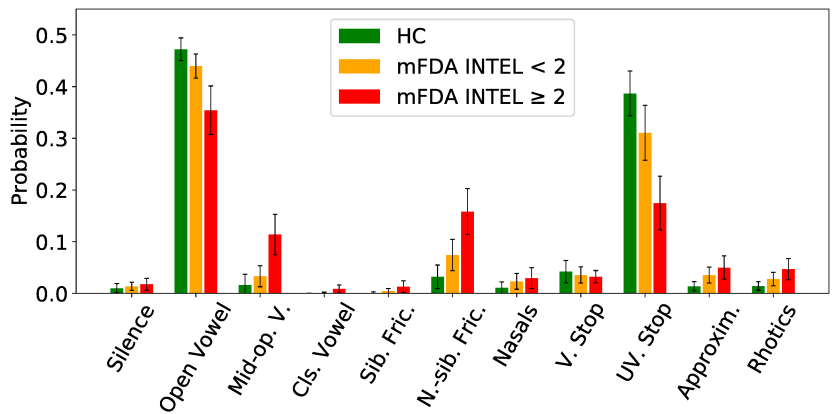}
  \caption{Mean posterior fine phonetic group probabilities for all DDK tasks.}
  \label{fig:ddk_fg}
\end{figure}

To better interpret the overall results, we visualized MPPs for the fine phonetic classes in Figure~\ref{fig:ddk_fg}. As we now looked at the complete set of exercises, we did not sort by an mFDA item associated to one particular articulator (e.g. the lips). Instead, we split the PD group into two sets by their overall intelligibility score. Throughout all exercises, we observed that the MPP for open vowels decreased along with the severity of speech symptoms experienced by a patient. In that case, it was very common for individuals to replace [a] with schwa [\textschwa] which was included in \textsc{phone} class [e]. The slight decline of MPP for voiced stops and the increasing amount of intermittent silence were both not considered significant with a t-test: for \mbox{silence: p = 0.11}, for voiced \mbox{stop: p = 0.06}. On the other side, the increasing probabilities of non-sibilant fricatives and the decline for unvoiced stops were found to be more expressive.\\
Table~\ref{tab:stats_ddk} shows the results for a statistical evaluation of the entire DDK results. We found a positive correlation with mFDA intelligibility scores for MPPs of mid-open vowels and non-sibilant fricatives and a negative correlation for the open vowels. A clear negative correlation was found for the unvoiced stop. Effect sizes were large for the set of PD patients with an intelligibility score below 2 for three out of the four tasks. For the set of poorly intelligible patients, we observed strong effect sizes compared to the healthy reference.

\begin{table}[t]
\small
\centering
\caption{Correlation between phonetic classes' mean posterior probability and mFDA intelligibility score. Columns 3 and 4 show effect sizes (Cohen's \textit{d}) between healthy reference and the respective split according to mFDA intelligibility score.}
\label{tab:stats_ddk}
\begin{tabular}{l|p{6em}|p{8em}|p{8em}}
 \cellcolor[HTML]{C0C0C0}Phonetic class & \cellcolor[HTML]{C0C0C0}Spearman's $\rho$ & \cellcolor[HTML]{C0C0C0}mFDA Intel. $<$ 2 & \cellcolor[HTML]{C0C0C0}mFDA Intel. $\geq$ 2\\ \hline
Mid-open Vowel	&	0.57	&	0.41   &   1.81 	\\ \hline
\rowcolor[HTML]{E8E8E8}Open Vowel   &   -0.59	&	0.72   &   1.92	\\ \hline
Non-sib. Fricative	&	0.64	&	0.79   &   2.11 	\\ \hline
\rowcolor[HTML]{E8E8E8}Unvoiced Stop  	& -0.71	&	0.79   &   2.32 	\\
\end{tabular}
\end{table}

\begin{figure}[t]
  \centering
  \includegraphics[width=0.9\linewidth]{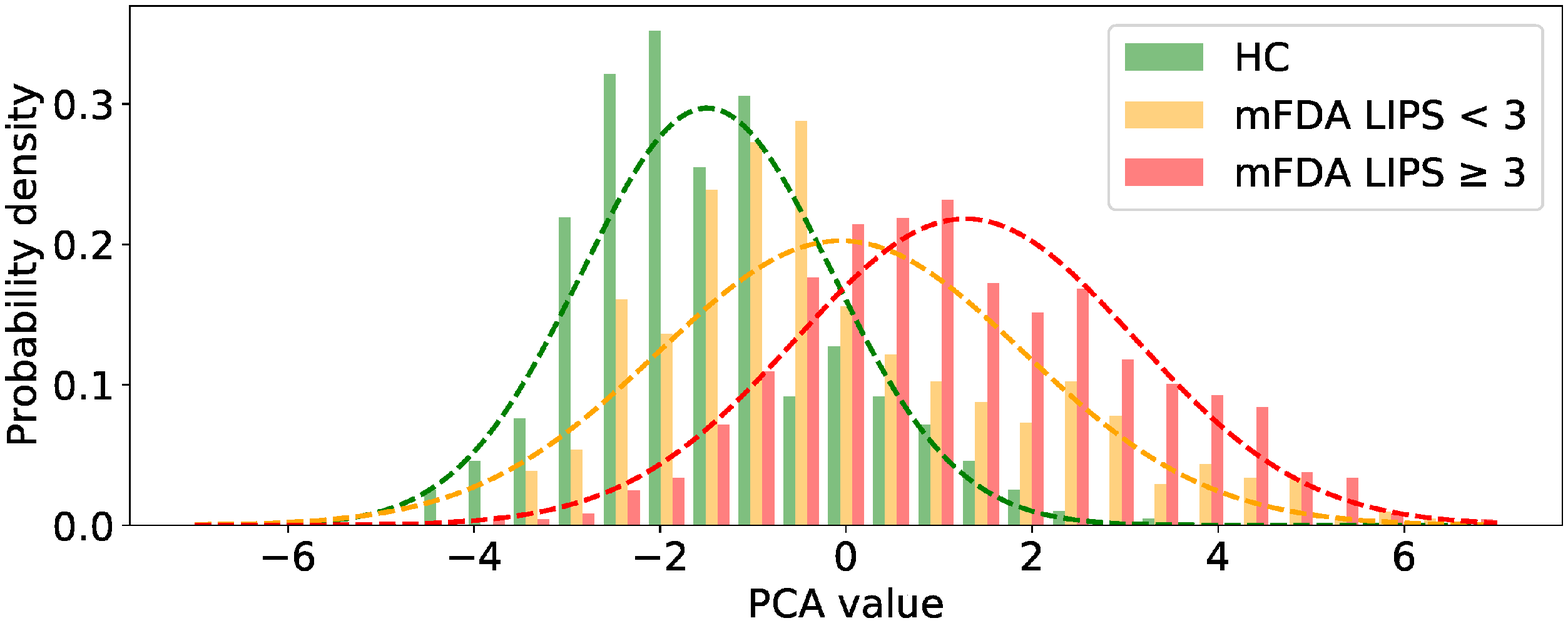}
  \caption{Distribution of PCA values computed from RNN hidden states at stop [p] (syllable [pa]) for HC, mildly and severly affected PD patients according to their mFDA lips score. Dashed curves illustrate underlying Gaussian distributions.}
  \label{fig:ddk4_pca}
\end{figure}

\begin{figure}[t]
  \centering
  \includegraphics[width=0.9\linewidth]{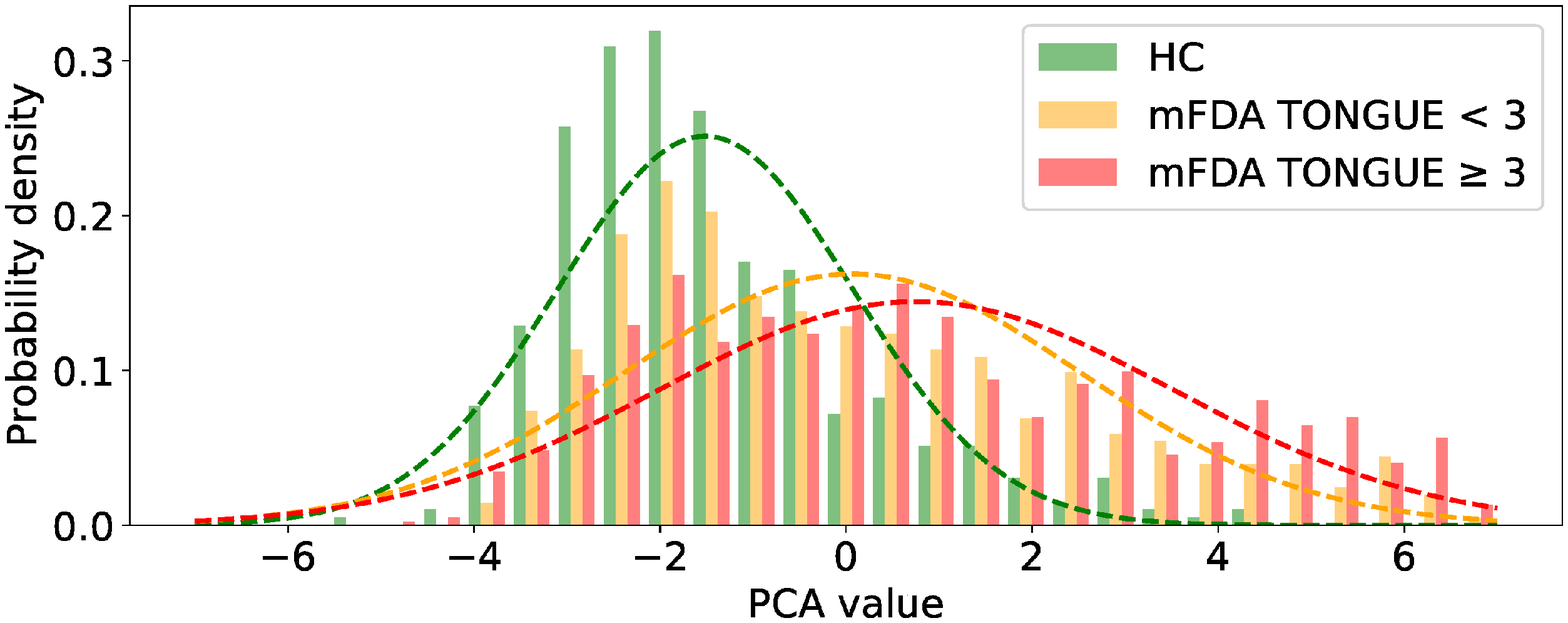}
  \caption{Distribution of PCA values computed from RNN hidden states at stop [t] (syllable [ta]) for HC, mildly and severly affected PD patients according to their mFDA tongue score. Dashed curves illustrate underlying Gaussian distributions.}
  \label{fig:ddk5_pca}
\end{figure}

\begin{figure}[t]
  \centering
  \includegraphics[width=0.9\linewidth]{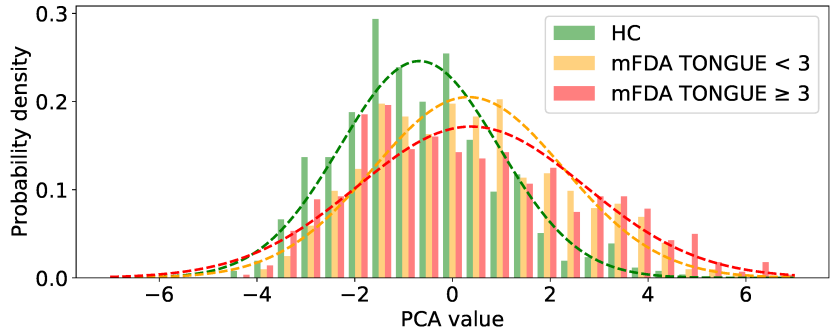}
  \caption{Distribution of PCA values computed from RNN hidden states at stop [k] (syllable [ka]) for HC, mildly and severly affected PD patients according to their mFDA tongue score. Dashed curves illustrate underlying Gaussian distributions.}
  \label{fig:ddk6_pca}
\end{figure}

\subsubsection{Intermediate network activations}
The PCA results for the three DDK tasks are depicted in Figures~\ref{fig:ddk4_pca}, \ref{fig:ddk5_pca} and \ref{fig:ddk6_pca}. For the syllable [pa], the mean PCA value of the HC group was $-1.5$ ($\sigma^2$~=~1.3). Mean values of the PD patients with low and high mFDA lips score were $-0.06$ ($\sigma^2$~=~2.0) and 1.3 ($\sigma^2$~=~1.8), respectively. We observed a marked deviation in the PCA embeddings of hidden states between the three sets, although all hidden states were ultimately classified to the same \textsc{phone}, the unvoiced stop [p].\\
The results for syllable [ta] indicated that the production of the stop sound [t] would also change for PD patients. Here, the mean PCA value of the control group was $-1.5$ ($\sigma^2$~=~1.6). Patients with an mFDA tongue score of not more than 2 had an average PCA value of 0.1 ($\sigma^2$~=~2.5), and those with scores above 2 were found to have $\mu=0.8$ ($\sigma^2$~=~2.8). Compared to the results of the first DDK task, the PCA values of the two PD sets showed a greater overlap for syllable [ta].\\
The last DDK task to evaluate was the production of stop sound [k] in the syllable [ka]. Here, we found the smallest deviation between PCA values among the different sets when sorting them according to the mFDA tongue score. The mean value of the control group was $-0.7$ ($\sigma^2$~=~1.6). The PD patients with low and high mFDA scores showed mean values of 0.3 ($\sigma^2$~=~1.9) and 0.4 ($\sigma^2$~=~2.3).\\
We also sorted the PCA results by mFDA intelligibility score, which resulted in greater differences between the sets. This was particularly interesting for the stops [t] and [k], because we already found [p] stops to show a good separation when sorted according to the involved articulator's score (lips). The mean values for [t] of the three sets after arranging them by intelligibility were $-1.5$ ($\sigma^2$~=~1.6), 0.4 ($\sigma^2$~=~2.7) and 1.0 ($\sigma^2$~=~2.0). For the last DDK task and stop [k], we observed mean values of $-0.7$ ($\sigma^2$~=~1.6), 0.2 ($\sigma^2$~=~2.1) and 1.4 ($\sigma^2$~=~2.2).

\begin{figure}[t]
  \centering
  \includegraphics[width=0.8\linewidth]{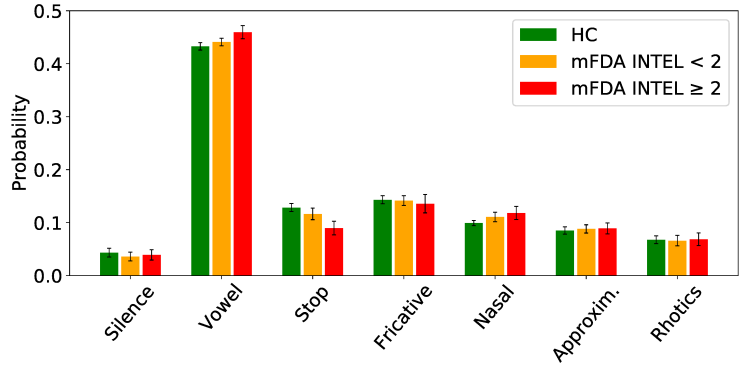}
  \caption{Mean posterior probabilities of coarse phonetic classes computed for isolated sentences.}
  \label{fig:sentence_cg}
\end{figure}

\begin{figure}[t]
  \centering
  \includegraphics[width=0.8\linewidth]{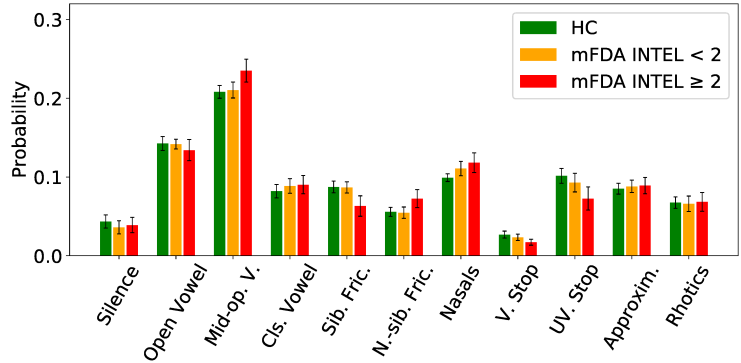}
  \caption{Mean posterior probabilities of fine phonetic classes computed for isolated sentences.}
  \label{fig:sentence_fg}
\end{figure}

\subsubsection{Isolated sentences}
In the results for isolated sentences, we rediscovered some of the patterns observed for the dedicated speech exercises already. Figures~\ref{fig:sentence_cg} and \ref{fig:sentence_fg} show the MPPs of coarse and fine phonetic classes. The most pronounced change between sets sorted by mFDA intelligibility score was observed for the stop sounds of the coarse phonetic categories with values of 12.8\% for the HC, 11.7\% for mildly and 9.0\% for severely affected PD patients. This pattern was identical to the trends observed for the DDK tasks, and once more we found the results to be significant (p-value $\ll$ 0.001). The increase of mean posterior probability of fricative sounds could not be found here. However, we found a pronounced shift of MPP from sibilant to non-sibilant fricatives for the poorly intelligible PD group in the fine phonetic categories. We also noticed an increasing MPP for nasal \textsc{phones}, decreasing open and increasing mid-open vowels.

\begin{figure}[t]
  \centering
  \includegraphics[width=0.8\linewidth]{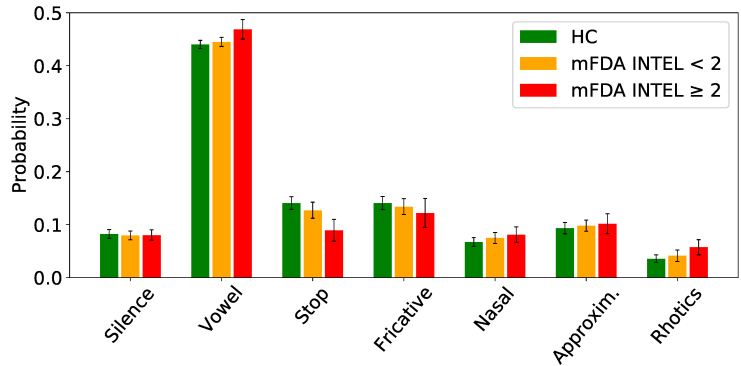}
  \caption{Mean posterior probabilities of coarse phonetic classes computed for the text reading task.}
  \label{fig:readtext_cg}
\end{figure}

\begin{figure}[t]
  \centering
  \includegraphics[width=0.8\linewidth]{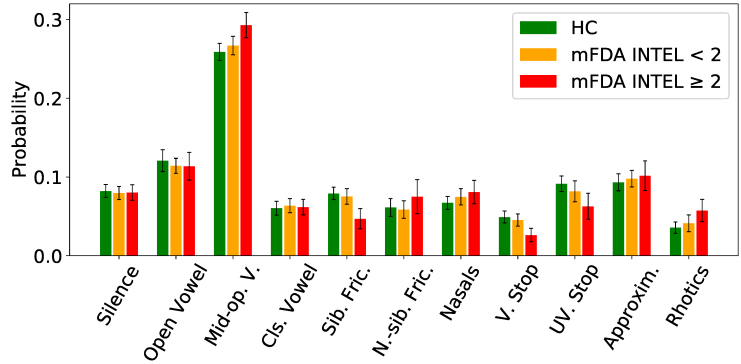}
  \caption{Mean posterior probabilities of fine phonetic classes computed for the text reading task.}
  \label{fig:readtext_fg}
\end{figure}

\subsubsection{Text reading}
For the task of reading a short text, the decreasing MPP of stop sounds was found once more in the coarse phonetic classes (Figure~\ref{fig:readtext_cg}). This time, the difference was a bit more pronounced with values of 14.1\% for the HC, 12.7\% for the mildly affected and 8.9\% for the severely affected PD patients. We also checked the nasal sounds in this scenario and found differences with values of 6.7\%, 7.5\% and 8.1\%. The increasing MPP of the rhotics was also noticeable, but we did not find this pattern in the isolated sentences. Looking at the results for the fine phonetic categories in Figure~\ref{fig:readtext_fg}, we rediscovered the decreasing amount of sibilant and increasing ratio of non-sibilant fricatives for the poorly intelligible PD participants that we already observed before. The increasing MPP of mid-open vowels was also preserved.

\begin{figure}[t]
  \centering
  \includegraphics[width=0.8\linewidth]{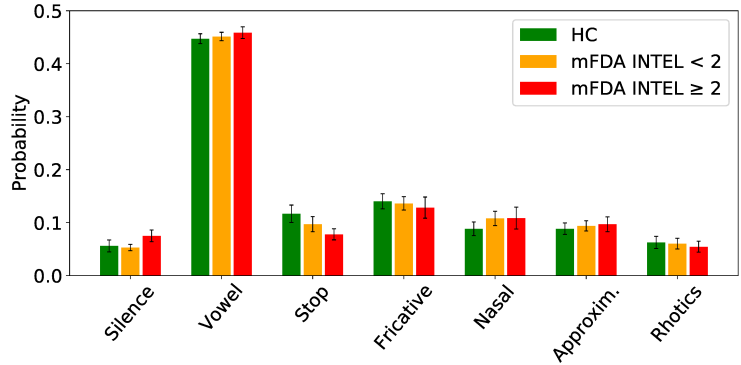}
  \caption{Mean posterior probabilities of coarse phonetic classes computed for the monologue.}
  \label{fig:monologue_cg}
\end{figure}

\begin{figure}[t]
  \centering
  \includegraphics[width=0.8\linewidth]{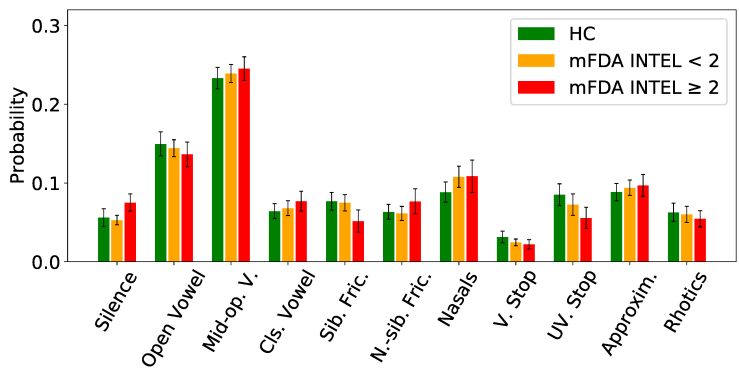}
  \caption{Mean posterior probabilities of fine phonetic classes computed for the monologue.}
  \label{fig:monologue_fg}
\end{figure}

\subsubsection{Monologues}
The final phonetic analysis results were computed for the monologue where the patients produced free speech. Results are shown in Figures~\ref{fig:monologue_cg} and \ref{fig:monologue_fg}. The most prominent differences in the coarse categories was again the decline of MPP for stop \textsc{phones}. The HC had an MPP of 11.7\%, this decreased to 9.7\% for the mild PD set and 7.8\% for the severe set. The latter equaled to a relative reduction of 33\% compared to the HC group. We also found an increased amount of nasal sounds for both PD sets, as we did for the previous tasks. An interesting result was the marked increase in MPP for silent segments (7.6\% compared to 5.6\% and 5.3\%) found for the poorly intelligible PD patients. We did not observe any such pattern in the previous tasks. The increasing MPP of trill \textsc{phones} found for the text reading task was not corroborated in the monologues. In the fine \textsc{phone} categories, we found the substitution of open vowels with their mid-open counterparts as we did for the other tasks. Similarly, the increasing MPP of the poorly intelligible PD group with respect to non-sibilant fricatives was conserved.

\begin{table}[t]
\small
\centering
\caption{Correlation between unvoiced stop (fine groups) mean posterior probabilities for different speech tasks and mFDA intelligibility score. Columns 3 and 4 show effect sizes (Cohen's~\textit{d}) between healthy reference and the respective split according to mFDA intelligibility score.}
\label{tab:stats_all_uvstop}
\begin{tabular}{l|p{6em}|p{8em}|p{8em}}
 \cellcolor[HTML]{C0C0C0}Speech task & \cellcolor[HTML]{C0C0C0}Spearman's $\rho$ & \cellcolor[HTML]{C0C0C0}mFDA Intel. $<$ 2 & \cellcolor[HTML]{C0C0C0}mFDA Intel. $\geq$ 2\\ \hline
Sentences	&	-0.50	&	0.40   &   1.30 	\\ \hline
\rowcolor[HTML]{E8E8E8}Text reading   &   -0.36	&	0.41   &   1.19	\\ \hline
Monologue	&	-0.43	&	0.46   &   1.08 	\\
\end{tabular}
\end{table}

\begin{table}[t]
\small
\centering
\caption{Correlation between stop (coarse groups) mean posterior probabilities for different speech tasks and mFDA intelligibility score. Columns 3 and 4 show effect sizes (Cohen's~\textit{d}) between healthy reference and the respective split according to mFDA intelligibility score.}
\label{tab:stats_all_stop}
\begin{tabular}{l|p{6em}|p{8em}|p{8em}}
 \cellcolor[HTML]{C0C0C0}Speech task & \cellcolor[HTML]{C0C0C0}Spearman's $\rho$ & \cellcolor[HTML]{C0C0C0}mFDA Intel. $<$ 2 & \cellcolor[HTML]{C0C0C0}mFDA Intel. $\geq$ 2\\ \hline
Sentences	&	-0.58	&	0.65   &   2.07 	\\ \hline
\rowcolor[HTML]{E8E8E8}Text reading   &   -0.45	&	0.50   &   1.74	\\ \hline
Monologue	&	-0.48	&	0.63   &   1.29 	\\
\end{tabular}
\end{table}

\subsubsection{Statistical interpretation}
Tables~\ref{tab:stats_all_uvstop} and \ref{tab:stats_all_stop} show the detailed evaluation of results for the most affect fine (Unvoiced stop) and coarse (Stop) phonetic classes. Throughout all tasks, we observed medium negative correlations between MPPs of the groups and the mFDA intelligibility score. This correlation was slightly stronger for the coarse groups. We observed large effect sizes for strongly affected patients. For the mildly affected group, these values are considerably lower.

\begin{table}[t]
\small
\centering
\caption{Mean confidence (posterior probability) in percent associated with a recognized \textsc{phone}. PD sets were arranged according to total mFDA score.}
\label{tab:confidences}
\begin{tabular}{l|p{6em}|p{6em}|p{6em}|p{6em}}
 & \cellcolor[HTML]{C0C0C0}HC & \cellcolor[HTML]{C0C0C0}mFDA < 20 & \cellcolor[HTML]{C0C0C0}mFDA < 30 & \cellcolor[HTML]{C0C0C0}mFDA $\geq$ 30 \\ \hline
DDK tasks	&	87.3	&	84.8   &   83.6   &   79.0	\\ \hline
\rowcolor[HTML]{E8E8E8}Sentences   &   86.2	&	83.8   &   82.4   &   78.0	\\ \hline
Read text	&	85.7	&	82.8   &   82.0   &   77.7	\\ \hline
\rowcolor[HTML]{E8E8E8}Monologue  	&	85.1	&	82.3   &   81.0   &   76.7	\\
\end{tabular}
\end{table}

\subsubsection{Prediction confidence}
We also evaluated the confidence of recognition results. To do so, we computed the mean posterior probability over all \textsc{phone} predictions for the different tasks. Results are summarized in Table~\ref{tab:confidences}. For each of the tasks, we observed a decrease in confidence as the total mFDA scores of patients increased. We found this decreasing confidence to be consistent throughout all tasks. Additionally, we found that the more freedom we had in our tasks (going from very restricted speech exercises to completely free speech), the overall confidence slightly decreased. It is important to notice that even for the free speech task (monologue), the mean posterior probability for the controls was higher than that of the mildly affected patients during the DDK exercises.

\begin{table}[t]
\small
\centering
\caption{Correlation between mean \textit{phone} confidences for different speech tasks and total mFDA score. Columns 3, 4 and 5 show effect sizes (Cohen's~\textit{d}) between healthy reference and the respective split according to total mFDA score.}
\label{tab:stats_conv_total}
\begin{tabular}{l|p{6em}|p{6em}|p{6em}|p{6em}}
 \cellcolor[HTML]{C0C0C0}Speech task & \cellcolor[HTML]{C0C0C0}Spearman's $\rho$ & \cellcolor[HTML]{C0C0C0}mFDA $<$ 20 &
 \cellcolor[HTML]{C0C0C0}mFDA $<$ 30 & \cellcolor[HTML]{C0C0C0}mFDA $\geq$ 30\\ \hline
DDK	&	-0.53	&	0.49   &   0.68 & 1.75	\\ \hline
\rowcolor[HTML]{E8E8E8}Sentences   &   -0.64	&	0.63   &   1.01 & 2.12	\\ \hline
Read text	&	-0.65	&	0.79   &   1.00 & 2.22	\\ \hline
\rowcolor[HTML]{E8E8E8}Monologue   &   -0.67	&	0.78   &   1.07 & 2.40	\\
\end{tabular}
\end{table}

\begin{table}[t]
\small
\centering
\caption{Correlation between mean \textit{phone} confidences for different speech tasks and mFDA intelligibility score. Columns 3 and 4 show effect sizes (Cohen's~\textit{d}) between healthy reference and the respective split according to mFDA intelligibility score.}
\label{tab:stats_conv_intel}
\begin{tabular}{l|p{6em}|p{8em}|p{8em}}
 \cellcolor[HTML]{C0C0C0}Speech task & \cellcolor[HTML]{C0C0C0}Spearman's $\rho$ & \cellcolor[HTML]{C0C0C0}mFDA Intel. $<$ 2 & \cellcolor[HTML]{C0C0C0}mFDA Intel. $\geq$ 2\\ \hline
DDK	&	-0.63	&	0.58   &   1.95 	\\ \hline
\rowcolor[HTML]{E8E8E8}Sentences   &   -0.74	&	0.74   &   2.07	\\ \hline
Read text	&	-0.70	&	0.81   &   2.43 	\\ \hline
\rowcolor[HTML]{E8E8E8}Monologue   &   -0.76	&	0.97   &   2.79	\\
\end{tabular}
\end{table}

Tables~\ref{tab:stats_conv_total} and \ref{tab:stats_conv_intel} provide the results of an evaluation of findings for the confidence scores. Mean confidence values showed medium to strong correlations to the total mFDA score. Notice that the correlations and effect sizes increased along with increasingly complex speech tasks. Correlations were found to be even more pronounced when compared to only the mFDA intelligibility item, with the highest (absolute) correlation of $-0.76$. Effect sizes showed clear differences between sets once again. As we already observed for the total mFDA, the lowest correlation of mean confidence with mFDA intelligibility was observed for the DDK task.

\section{Discussion}
\subsection{Speech exercises}
The results for all DDK speech exercises showed clear differences between the HC group and the PD patients. This difference was also observed to markedly increase along with the mFDA scores associated with the involved articulators. We also found the results to be well interpretable. A deficit in muscle tension lead to slurred pronunciation of stop \textsc{phones}, where it was increasingly challenging for patients to produce the characteristic closure-burst combination of unvoiced stops. Without the full closure of the lips for example, a [p] would transition into an [f] or [v]. In this context, the transition to [v] is particularly important, because in Colombian Spanish there is no acoustic/phonetic difference between [p] and [v]. The difficulties related to lip muscle control observed from the phonetic profile for stop sound [p] are well-known~\cite{leanderson1972lip}. Similar patterns were found for the other places of articulation and their respective \textsc{phones} [t] (substitution with [\dh] (voiced English \textit{th}) or [l]) and [k] (substitution with [r], [\textgamma] or [x]). For syllable [ta], the observation that mildly affected PD patients were on average more likely to correctly produce [t] was surprising. The high MPP of healthy speakers for the stop [d] could give serve as an explanation that some speakers in the HC group tended to produce more voiced [d] than unvoiced stops [t]. Overall, they were still more likely to produce a stop than any PD set Notice that the position of the involved articulators (lips or tongue) and the manner of articulation of the different stop sounds were preserved in the transition \textsc{phones}, but the required muscle tension was lost, thus making it difficult to correctly produce the target \textsc{phone}. Throughout all DDK tasks, we found that the PD patients showed an increased instability for producing the vowel [a]. This instability was already found to be significant in other studies \cite{sakar2013collection,tsanas2012novel}. Instead, patients produced more mid-open (centralized schwa-like) vowels and showed a reduced ability to fully open their vocal tract, possibly due to increasing rigidity~\cite{Proenca2013}. In general, for all exercises of stop-vowel syllable repetitions, we found that PD patients showed clear difficulties to produce stop sounds reliably and instead were more likely to replace them with fricatives.\\
The strong negative correlation of unvoiced stop \textit{phone} MPPs and the mFDA intelligibility score clearly indicated that an increasing effect of PD on an individual's speech would result in a reduced MPP of unvoiced stop sounds. On the other hand, non-sibilant fricatives turned out to be positively correlated to the score. This could be caused for example by a transition from stop \textit{phones} or sibilant fricatives over to the non-sibilants, but this is subject of further investigation.\\
The high standard deviations observed for speech exercises in stop sounds and fricatives were likely caused by confusions of unvoiced and voiced stops (both during production and recognition) as well as spontaneous transitions from stops to fricatives.
\subsection{Intermediate network activations}
With the PCA decomposition of hidden states that were classified as the correct stop \textsc{phones} [p], [t] or [k], we tried to find if there were noticeable differences in \textsc{phone} productions between HC and PD, even if the correct \textsc{phone} was detected. Not only did our results indicate these differences, but they also became more pronounced with increasing mFDA score of the involved articulator. Because the only objective of a PCA decomposition is to preserve the maximum amount of variation from a higher-dimensional feature space, we assume that the disease state introduced such a variation in the hidden states of stop sound productions and that it was sufficiently large to remain visible after the one-dimensional PCA. We noticed the largest discrepancy between sets for the stop [p] if we arranged sets by mFDA lips rating. This seemed plausible because the lips play the major role in creating the obstruction for the closure and release for the burst of the corresponding stop \textsc{phone}. Similar discoveries were made for the stops [t] and [k], but there, the difference between mildly and strongly affected patients was not as big. For \textsc{phone} [k], it is likely that the tongue item was also not perfectly resembling a patient's ability to produce velar \textsc{phones}. This assumption was further supported by the fact that the observed differences between the sets were higher for [t] and [k] when we sorted by mFDA intelligibility instead of mFDA tongue rating. We chose a subset of the PD speech corpus described in section~\ref{subsec:PDdata} to ensure equal acoustic recording conditions for HC and PD subjects. Therefore, we could attribute the observed differences in the hidden states to the effects of PD on an individual's speech production.
\subsection{Isolated sentences and text reading}
Whilst the differences in MPP for the stop sounds were very pronounced in the speech exercises, they were expected to lower when analysing spoken sentences or short texts. Nevertheless, we found that for both tasks, the MPP of unvoiced and voiced stops continued to decline along with the rated intelligibility. At the same time, we observed a new pattern, an increasing amount of nasal sounds. The development of a measurable hypernasality caused by PD has been reported previously~\cite{novotny2016hypernasality}. It is likely caused by reduced control over the velopharyngeal tract due to loss of muscle tension, thereby allowing air to pass through the nasal cavity~\cite{godino2017towards}. Our results indicate that this hypernasality could be found with the presented \textsc{phone} recognizer, but it is yet to be confirmed that the nasalization of vowels would result in an increased MPP of nasal \textsc{phones}. As for the speech exercises, PD patients were more likely to produce fewer open and more mid-open vowels as the disease progressed. An impaired vowel articulation was observed in related works~\cite{skodda2012impairment,whitfield2014articulatory}, which resulted in a smaller vowel space and would therefore lead to fewer realizations of open vowels.  Our findings for isolated sentences and read texts among different sets supported the assumption that an impaired speech production and the linked decline in intelligibility would leave a detectable footprint in the phonetic profile of affected individuals. With respect to the correlation between MPPs and intelligibility scores, it did not make much difference if we arranged \textit{phones} into fine (unvoiced stop) or coarse (stop) classes. In both cases, we observed medium correlations for spoken sentences and text reading exercises.
\subsection{Monologues}
One key research question was if it would be possible to translate any of the aforementioned findings to free speech. The results from speech recordings of monologues indicated that many of the patterns observed in the previous experiments were present in free speech as well. For example, the deterioration of stop sounds, the increasing amount of nasality and the transition from open to mid-open vowels were conserved. The previously described articulation deficits (reduced vowel space, impaired control over velopharyngeal port and overall decreased muscle tension) manifest in the phonetic footprints of PD patients during free speech. MPPs of (unvoiced) stop sounds once more showed a medium correlation to mFDA intelligibility scores in the monologue. Another important finding was the ratio of silent segments for the poorly intelligible PD patients. Relative to the mildly affected PD group and the HC (which showed only minor differences), they showed 39\% more silent events. These intermittent pauses could potentially be related to a mild cognitive impairment (MCI) often developing in the later stages of PD. MCI in PD has already been described in previous studies~\cite{verbaan2007cognitive,kehagia2010neuropsychological,aarsland2010mild}. Speech disfluencies in PD have been investigated in detail before~\cite{goberman2010characteristics} and they could be attributed to two major causes, stuttering and hesitations. Stuttering occurred more frequently within words and was provoked by mispronunciations. Hesitations were found more frequently between words and were far more likely to happen in free speech tasks. With our \textsc{phone} recognizer, we were able to find patterns of hesitations as well for poorly intelligible PD patients, and we could confirm their predominant occurrence in free speech.
\subsection{Prediction confidence}
As our \textsc{phone} recognition model was trained only with healthy speech, we also tried to investigate if impaired speech samples would result in an overall decreased confidence of network predictions. Throughout all tasks, we observed high confidences for the HC group (minimum of 85.1\,\% for monologues). To put this into more context, the mean confidence computed over the test set of all three languages of our training data was almost 90\%. The control group was very close to this value. With increasing mFDA scores of our PD sets, we found the recognizer mean confidence to decline equally for every task. The slight decline within every set going from dedicated speech tasks to free speech was expected due to the increasing variability and length of the signals. The fact that the mean confidence for monologues of HC speakers was still higher than that of any other PD task or set lead us to the conclusion that our model had a decent generalization of what healthy speech sounded like. Accordingly, predictions became markedly less confident as the severity of PD increased. Confidences were clearly correlated to both mFDA total and intelligibility scores. These correlations were particularly pronounced in the monologues and lowest for the dedicated speech exercises. A possible explanation is the rather short duration of an exercise (a couple of seconds) compared to that of a monologue (a few minutes). The longer a patient had to speak, the more likely they were to show signs of a declining articulation. Compared to the findings reported in \cite{van2009speech}, their models showed stronger correlations (0.793 to 0.943) with intelligibility scores. However, they employed dedicated regression models for score prediction and did not correlate to individual features such as prediction confidences, as it was done here. Furthermore, our models do not require any transcription and yield promising results even for free speech.

\section{Conclusion}
We presented a neural network for \textsc{phone} sequence recognition trained on datasets from three different languages. Pretraining the network with alignment information helped to increase the accuracy of the final CTC model. We then used the recognizer to analyze speech samples from a PD dataset and created phonetic profiles for different subgroups over various tasks. For the dedicated speech exercises, our \textsc{phone} recognition model was able to find clear differences in the mean posterior probabilities of stop sounds, which were also strongly correlated to mFDA intelligibility scores, and instabilities in vowel productions of PD patients. Furthermore, it was possible to transfer these findings to more complex speech samples. Even in free speech recordings, the decreased amount of stop sounds was observed. We also found other patterns that have previously been reported in other studies, such as increased nasalization, reduced vowel articulation space or, for the severely affected PD patients, an increased amount of pauses. Particularly for the nasalization, it could be interesting to incorporate a French healthy speech dataset to include nasalized vowels in the \textsc{phone} space.\\
With the intermediate hidden state vectors from the RNN, we showed that even though all observed hidden state were classified to the same stop sound, they showed clear differences after applying a PCA decomposition that were directly correlated to the impairments of certain articulators.\\
The analysis of MPPs over all network predictions of a particular set showed that the network was most confident while predicting samples of the control group. Both the total mFDA score (and therefore the degree of impairment) as well as only the intelligibility item were clearly correlated to the mean confidences of \textsc{phone} predictions. This was explained by the network being trained exclusively on healthy speech. The less intelligible a speech signal sounded, the farther away it would be from the familiar hyperspace of healthy speech, thus resulting in a lower confidence.\\
In the field of pathological speech analysis, modern deep learning methods often lack the required amount of data to be applied to any such problem. Instead of training a model with pathological speech to identify characteristic patterns for classification (HC or PD) or regression (PD progression) tasks, it can be sufficient to train the entire model solely with large amounts of healthy speech. These models could serve as a reference of what healthy speech sounds like. At the same time, they enabled us to determine a phonetic footprint of a healthy cohort for a given language. Such footprints could also be computed for speaker groups of Parkinson's patients to compare their articulatory abilities to a healthy reference. Phonetic footprinting holds two major advantages over other methods. Models would not be prone to overfitting on a small domain-specific dataset, simply because they would never see such data in the process of parameter optimization. Secondly, the footprints and their differences between speaker groups are highly interpretable. Particularly in the field of pathologic datasets, this property is extremely important as it helps to better understand a classifier's decision process. In our case, we confirmed a number of speech patterns of PD patients that had been described in other studies before, although our model was trained on \textsc{phone} recognition and had never received any information about Parkinson's disease.

\section{Acknowledgements}
This study has been funded by the Federal Ministry of Education and Research of Germany in the ASA-KI project, grant No. 16SV8469. The authors also acknowledge to the Training Network on Automatic Processing of PAthological Speech (TAPAS), grant agreement No. 766287, and the EU project sustAGE, grant agreement No. 826506, both funded by the Horizon 2020 program of the European Commission. Tom{\'a}s Arias-Vergara is under grants of Convocatoria Doctorado \mbox{Nacional-785} financed by COLCIENCIAS. This work was also financed by CODI at UdeA grant No. PRG2017-15530.

\bibliography{elsarticle-template}
\newpage
\appendix
\section{Model architecture}
\label{appendix:arch}
\begin{figure}[t]
\centering
\begin{subfigure}[b]{0.4\textwidth}
  \centering
  \includegraphics[width=0.9\linewidth]{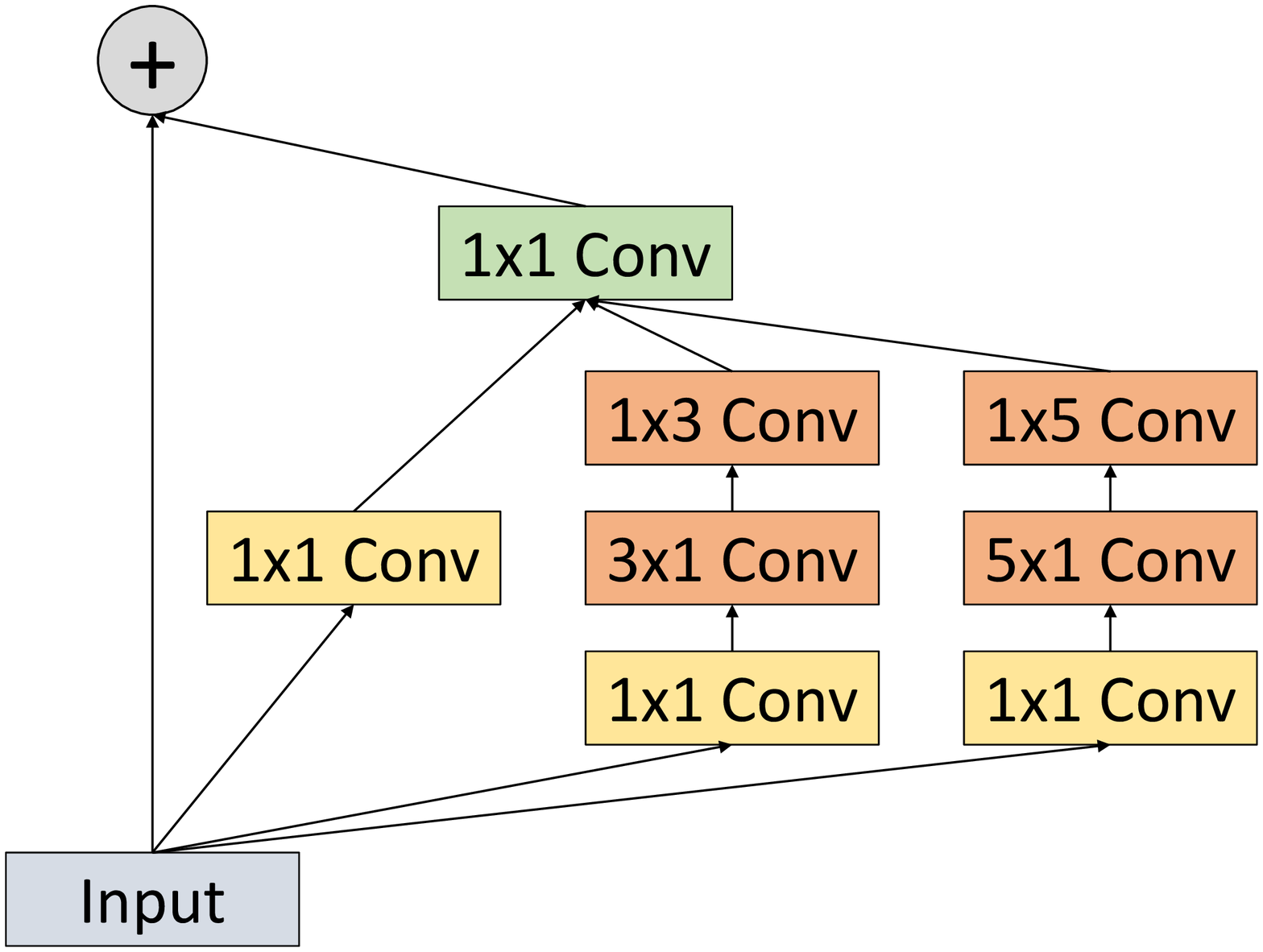}
  \caption{Residual inception block used in the \textsc{phone} recognition network.}
  \label{fig:res_inc_block}
\end{subfigure}
\hspace{1cm}
\begin{subfigure}[b]{0.4\textwidth}
  \centering
  \includegraphics[width=0.9\linewidth]{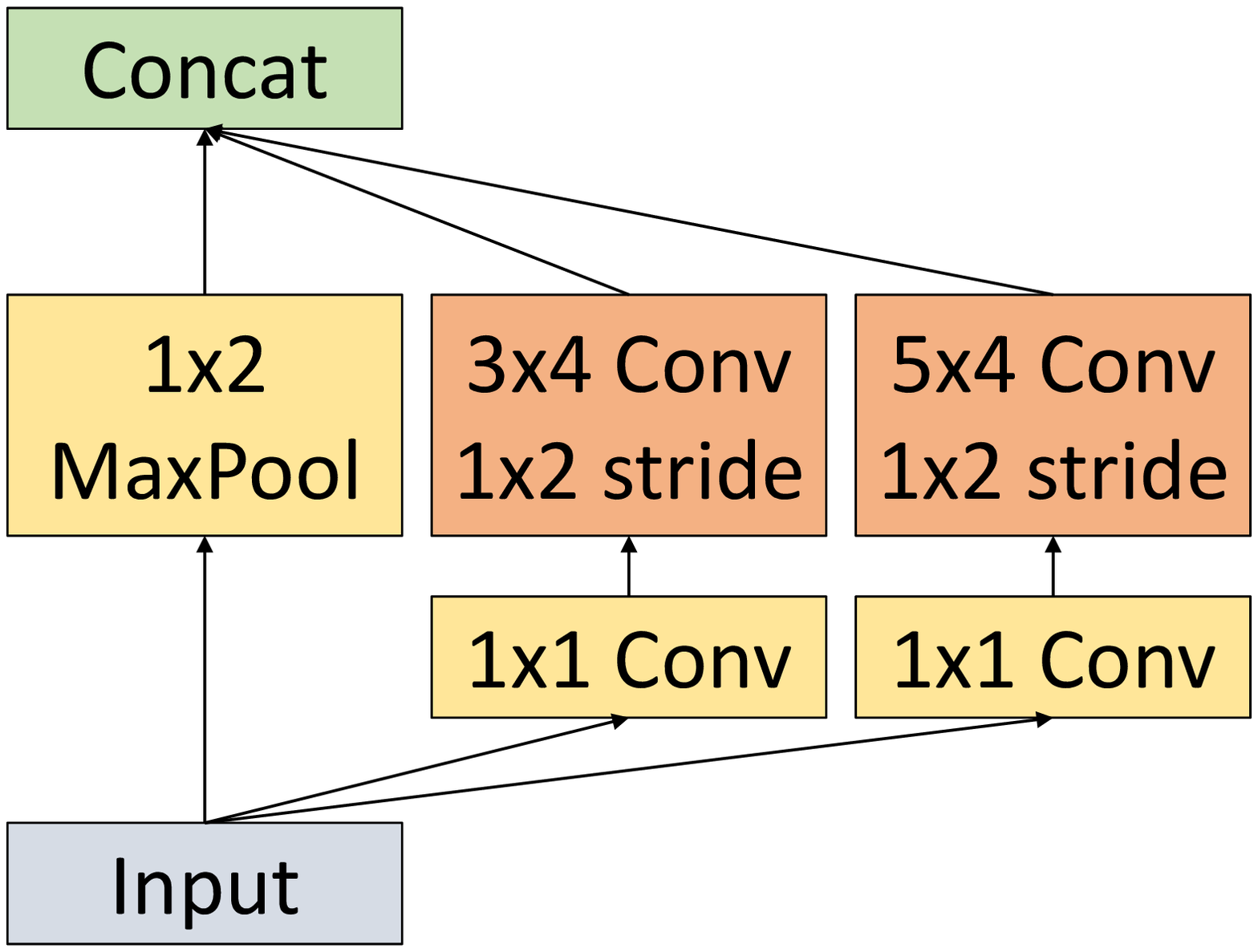}
  \caption{Reduction inception block used in the \textsc{phone} recognition network.}
  \label{fig:red_inc_block}
\end{subfigure}
\caption{Two types of inception blocks used in the \textsc{phone} recognition network}
\end{figure}

The convolutional feature extraction part was constructed with two major building blocks inspired by the Inception architectures presented for image classification~\cite{szegedy2017inception}. The core idea was to apply multiple convolution kernels of varying sizes in parallel such that the network itself would learn which kernel worked best for what task. Figure~\ref{fig:res_inc_block} depicts a residual inception block. The initial 1x1 convolutions perform a channel reduction to make the following convolution operations less parameter-intensive. The 3x3 and 5x5 convolutions which operated in parallel were realized with separated kernels to further reduce the required number of parameters. A final 1x1 convolution projected the concatenated results from the three branches back to the original number of channels. Before the output was then added to the input to realize a residual connection~\cite{he2016deep}, we applied activation scaling ($s=0.3$) as proposed in~\cite{szegedy2017inception} to stabilize training. The second important building block of our network was the reduction inception block shown in Figure~\ref{fig:red_inc_block}. The major purpose of that component was to reduce the remaining number of frequency bins while leaving the time dimension unchanged. Reduction was performed with parallel max-pooling and strided convolution layers and their outputs were concatenated.\\
The architectures of the CTC and the framewise model are described in Table~\ref{tab:archs}. In the first step, we used several 2D-convolutions to reduce the number of frequency bands and encode the corresponding information in the channel dimension. This part of a convolutional neural network (CNN) is commonly referred to as the stem. Note that the very first layer halved the number of time steps. Our dual-channel spectrograms were computed with a temporal resolution of 5~ms, which helped to reliably detect very short \textsc{phones}. For the final prediction however, a resolution of 10~ms was sufficient. After the stem, we used the residual and reduction inception blocks to learn feature maps and gradually reduce the number of remaining frequency bands. A depthwise separable convolution~\cite{sifre2014rigid} was used to project the four remaining frequency bands and their 580 channels after the last inception block down to 300 values per time step. The initial stem, the inception blocks and the depthwise separable convolution were considered the feature extraction part of the architecture. To perform a sequence analysis, we then applied a stack of two bidirectional Long Short-Term Memories (BiLSTM)~\cite{hochreiter1997long}. Each BiLSTM used a hidden state vector with 240 units in the case of CTC, 200 units during the pre-training. In the final step, the result for each time step was linearly projected to the number of targets, 35 for CTC, 44 for aligned pre-training. All hyperparameters, for example kernel sizes or channel configurations, were tuned for the best result in \textsc{phone} sequence prediction. Thus, the presented architecture was never optimized for PD speech analysis.

Hard swish~\cite{howard2019searching} was used as activation function. The outputs of the convolutional stem-layers, the residual inception blocks and the depthwise separable convolution were normalized through batch normalization~\cite{ioffe2015batch}. After each layer normalization, we applied a dropout of 10\% to prevent overfitting. The model for aligned pre-training comprised around 6.6 million parameters, the final CTC model was a bit larger with roughly 7.2 million parameters.

\begin{table}[t]
\small
\centering
\caption{Outline of the CTC \textsc{phone} recognition model. Output size depended on the length of the sample (\textit{T}). \#\textit{c} indicates number of channels. \#x\# denotes kernel size in temporal (first) and frequency (second) domain. [\#, \#] denotes the stride in the respective domain. For the recurrent neural network (RNN) layers and linear projection layers, numbers in brackets denote the configuration of the pretrained model with alignment information.}
\label{tab:archs}
\begin{tabular}{|c|c|}
\hline
\rowcolor[HTML]{C0C0C0} 
Output size & Layer                                                                                     \\ \hline
2Tx64, 60   & 60c 1x4 Conv {[}1, 2{]}                                                                   \\ \hline
Tx64, 120   & 120c 5x1 Conv {[}2, 1{]}                                                                   \\ \hline
Tx32, 160   & 160c 1x4 Conv {[}1, 2{]}                                                                   \\ \hline
Tx16, 200   & 200c 1x4 Conv {[}1, 2{]}                                                                   \\ \hline
Tx16, 200   & \makecell{2 x Residual Inception Block \\ ch. reduced: 70}        \\ \hline
Tx8, 340    & \makecell{Reduction Inception Block\\ ch. reduced: 70}       \\ \hline
Tx8, 340    & \makecell{2 x Residual Inception Block\\ ch. reduced: 120}        \\ \hline
Tx4, 580    & \makecell{Reduction Inception Block\\ ch. reduced: 120}       \\ \hline
Tx4, 580    & \makecell{2 x Residual Inception Block\\ ch. reduced: 200}        \\ \hline
Tx300       & Depthwise separable convolution                                                           \\ \hline
Tx480 (400) & 2 x BiLSTM 240 (200) hidden units                                                        \\ \hline
Tx35 (44)   & Linear projection to 35 (44) target \textsc{phones} \\ \hline
\end{tabular}
\end{table}

\end{document}